\documentclass[12pt]{article}
\pdfoutput=1
\usepackage{amsmath,amssymb,amsfonts,amsthm,bm,bbm,cancel,wasysym}
\usepackage{epsfig,graphics,graphicx,epstopdf,caption,subcaption}
\usepackage[numbers,sort&compress]{natbib}
\graphicspath{{Charts/}}
\usepackage{array,booktabs,colortbl,colordvi,multirow}
\usepackage{colordvi,color,xcolor}
\usepackage{hyperref}
\usepackage{rotating}
\usepackage{comment}
\usepackage{feynmf}

\usepackage[symbol]{footmisc}
\renewcommand{\thefootnote}{\fnsymbol{footnote}}

\begin{document}

\begin{center}

{\Large {\bf  Constraints on freeze-in dark matter from Lyman-$\alpha$ forest and 21-cm signal : single-field models}}\\
%\vspace*{0.15cm}

\vspace*{0.75cm}

{Zixuan Xu, Quan Zhou and Sibo Zheng\footnote{Corresponding author: sibozheng.zju@gmail.com}}

\vspace{0.5cm}
{Department of Physics, Chongqing University, Chongqing 401331, China}

\end{center}
\vspace{.5cm}

%--------------------------------------------- ABSTRACT ---------------------------------------------%
\begin{abstract}
\noindent 
We report new Lyman-$\alpha$ and 21-cm constraints on freeze-in dark matter (FIDM) which injects energy into the intergalactic medium either through annihilation or decay to photon(s) or electron-positron pair. 
With respect to Lyman-$\alpha$ we fix the baseline ionization history using low redshift data about astrophysical reionization,
whereas for 21-cm signal we adopt the baseline values of 21-cm power spectrum through a standard modeling of star formation developed so far.
Using the latest numerical tools, we show that (i) for sterile neutrino FIDM, current Lyman-$\alpha$ data and future sensitivity of SKA-low (1000 hrs) on the 21-cm power spectra excludes the FIDM mass up to $1.8\times 10^{-3}$ GeV at 95$\%$ CL and $5.46\times 10^{-4}$ GeV, respectively, 
and (ii) for millicharged FIDM, current Lyman-$\alpha$ data only excludes the millicharge down to $10^{-8}$ within the FIDM mass range of $10^{-3}-1$ GeV at 95$\%$ CL, suggesting that the surviving parameter space of millicharged FIDM is still intact.
\end{abstract}

\renewcommand{\thefootnote}{\arabic{footnote}}
\setcounter{footnote}{0}
\thispagestyle{empty}
\vfill
\newpage
\setcounter{page}{1}

\tableofcontents

\section{Introduction}
%%%  FIDM 
Because of null results of experiments aiming to detect dark matter (DM) as a weakly interacting massive particle (WIMP),
there is a renewed interest in freeze-in dark matter (FIDM) which differs from WIMP.  
Compared to WIMP-like DM, a FIDM is produced from the Standard Model (SM) thermal bath of the early Universe via so-called freeze-in mechanism \cite{Hall:2009bx},
as a result of FIDM feebly coupling to the SM sector.\footnote{To be concrete, we do not consider DM to freeze-in via inflaton sector after the end of inflation.}  
Despite being capable of addressing the observed DM relic density, 
such feeble coupling makes the FIDM be unlikely to leave observable footprints in the aforementioned experiments,
which asks for new detection strategies.

%%%% CMB constraints
In this work we consider cosmological probes of FIDM.
The early studies on effects of DM scattering \cite{Cyr-Racine:2015ihg} either off photons or baryons
provided Cosmic Microwave Background (CMB) constraints \cite{Dvorkin:2013cea,Xu:2018efh,Slatyer:2018aqg} being competitive with collider or direct detection limits for WIMP-like DM, which are however not viable for the FIDM. 
Instead of scattering, DM annihilation or decay into photons and/or electron-positron offers improved CMB constraints \cite{Padmanabhan:2005es,Slatyer:2009yq,Lopez-Honorez:2013cua,Diamanti:2013bia,Slatyer:2015kla,Poulin:2015pna,Bolliet:2020ofj,Dvorkin:2020xga, Capozzi:2023xie,Liu:2023nct, Li:2024xlr}, 
as a result of relevant observables having a lower power laws of the feeble coupling.
In this approach, 
the DM annihilation- or decay-induced energy injection into the intergalactic medium (IGM) modifies ionization history of baryon gas, leading to changes in CMB spectra.

%%%  improvements made by Ly-a and 21-cm 
Apart from the imprints in the CMB spectra, 
DM induced energy injection to the IGM also affects observations of Lyman-$\alpha$ \cite{Diamanti:2013bia,Capozzi:2023xie,Liu:2016cnk, Liu:2020wqz} and 21-cm signal \cite{Liu:2018uzy,DAmico:2018sxd,Mitridate:2018iag,Facchinetti:2023slb}.
Either photons or electron-positron arising from DM annihilation or decay 
deposit their energies in the IGM via heating, hydrogen ionization, helium single or double ionization and neutral atom excitation,
which changes the ionization history of IGM measured by Lyman-$\alpha$ and 21-cm experiments. 
Compared to the CMB constraints,
ref.\cite{Liu:2020wqz} shows that the Lyman-$\alpha$ lower (upper) bound on DM lifetime (annihilation cross section) can be improved by one-to-two (several) orders of magnitude in certain DM mass range for DM decay (annihilation) into $e\bar{e}$.
Similar improved ability of exclusion is also seen in the 21-cm constraints \cite{Facchinetti:2023slb}.
Note, the Lyman-$\alpha$ and 21-cm constraints rely on how to model astrophysical reionization and star formation respectively.

 %%% new strategy 
In this work we utilize observations of Lyman-$\alpha$ and 21-cm signal to place new constraints on explicit single-field FIDM models. 
Regarding the Lyman-$\alpha$ constraints, 
we use currently available data about the ionization parameters in low redshift region to infer the astrophysical reionization. 
Then we use publicly available numerical code to derive FIDM decay or annihilation induced deviations from the baseline values of these ionization parameters.
Requiring these deviations relative to the baseline values to be less than 2$\sigma$ order gives us the Lyman-$\alpha$ constraints at 95$\%$ CL.
For the 21-cm constraints, we follow a standard modeling of star formation developed so far,
which gives rise to the baseline values of 21-cm brightness temperature and power spectrum.
Likewise, we use publicly available numerical code to calculate FIDM induced deviations from the baseline values of 21-cm power spectrum.
Comparing these deviations to future sensitivities on the 21-cm power spectrum delivers the 21-cm constraints.

The rest of the paper is organized as follows. 
In Sec.\ref{models} we consider the explicit FIDM models via Higgs \cite{McDonald:2001vt, Kang:2015aqa}-, neutrino \cite{Asaka:2005cn,Becker:2018rve,Chianese:2018dsz,Datta:2021elq}- and vector \cite{Holdom:1985ag,Feldman:2007wj,Yin:2023jql}-portal respectively,
where we analyze each FIDM annihilation or decay induced energy injection into the IGM.
Sec.\ref{modeling} is devoted to model the FIDM induced effects on the evolution of IGM parameters and 21-cm observables,
where we will briefly introduce theoretical backgrounds and numerical tools used for later numerical analysis.  
In Sec.\ref{results} we report new Lyman-$\alpha$ and 21-cm constraints on the FIDM models studied in Sec.\ref{models}, 
where we will point out how these limits differ from the aforementioned results in the literature and compare them to existing bounds.
Finally, we conclude in Sec.\ref{con}.

%%%%%%%%%%%%%%%%%%%%%%%%%%%%
\section{Freeze-in dark matter induced energy injection into the IGM}
\label{models}
In this section we derive the DM induced energy injections into the IGM in three different single-field FIDM models.

\subsection{Higgs portal}
\label{h}
The first single-field FIDM is built upon the SM Higgs portal with the DM Lagrangian as \cite{McDonald:2001vt, Kang:2015aqa}
\begin{eqnarray}\label{HLag}
\mathcal{L}_{\rm{dark}}=\frac{1}{2}\partial_{\mu} X\partial^{\mu}X-\frac{\mu^{2}}{2}X^{2}-\kappa X^{2}\mid H\mid^{2}-\lambda X^{4},
\end{eqnarray}
where $X$ is the scalar DM, $H$ is the SM Higgs doublet, $\kappa$ is a small Yukawa coupling between the Higgs and DM, and $\lambda$ is the self-interacting DM coupling constant.
In eq.(\ref{HLag}) a hidden $Z_2$ parity, under which the DM is odd, has been assumed to make sure the completeness of $\mathcal{L}_{\rm{dark}}$.
We simply ignore the DM self-interaction, as it has no role to play in the following analysis.
In the broken electroweak phase the DM mass is given by $m^{2}_{X}=\mu^{2}+\kappa \upsilon^{2}$, with $\upsilon=246$ GeV the weak scale.
Therefore, the free parameters in this model are only composed of $\{m_{X}, \kappa\}$.

\begin{figure}
\centering
\includegraphics[width=8cm,height=8cm]{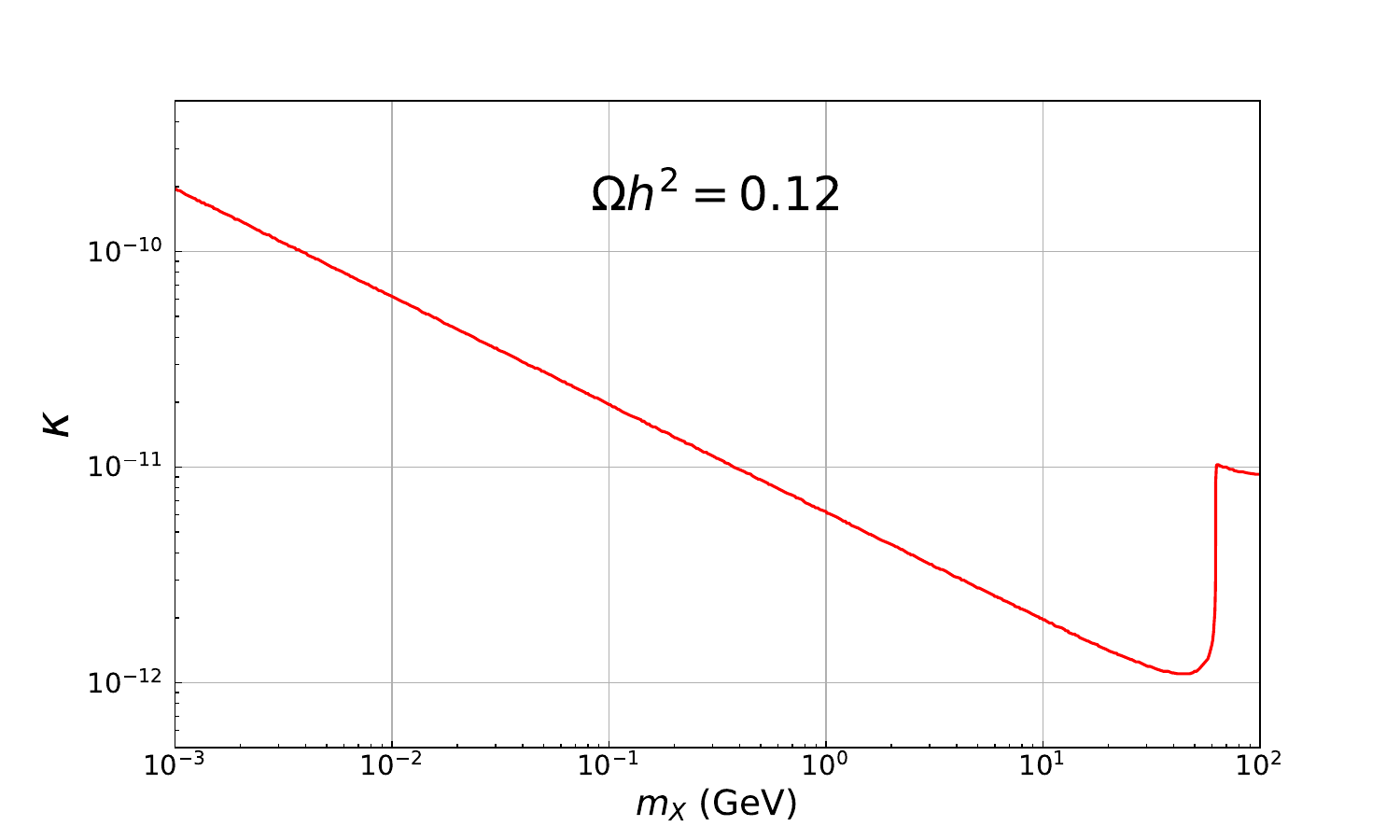}
\includegraphics[width=8cm,height=8cm]{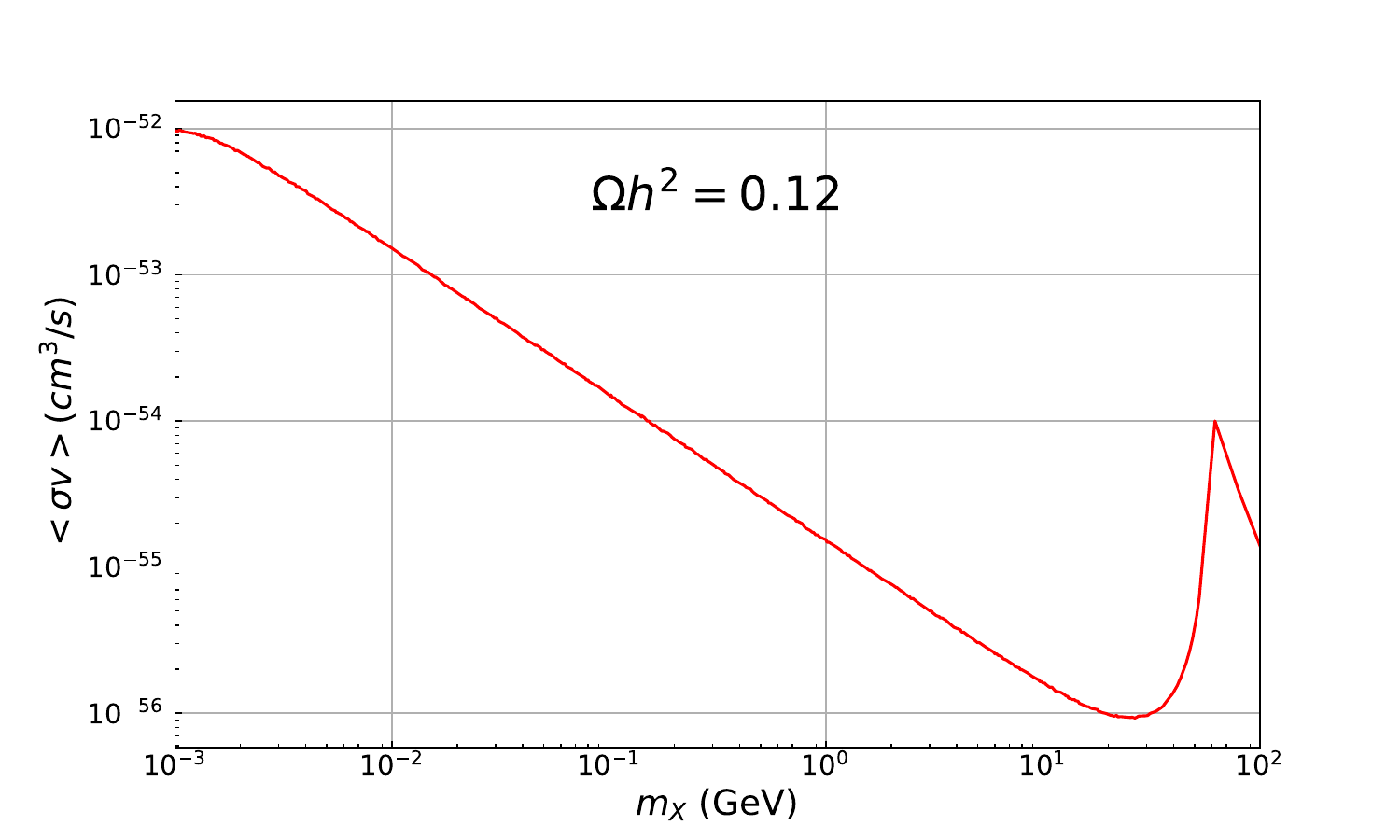}
\centering
\caption{$\mathbf{Left}$: the observed DM relic abundance $\Omega_{X}h^{2}=0.12\pm0.001$ projected to the plane of $m_{X}-\kappa$. $\mathbf{Right}$: the thermally averaged values of $\left<\sigma v(XX\rightarrow e\bar{e})\right>$, 
with the value of $\kappa$ fixed as in the left panel.}
\label{hdm}
\end{figure}

The left panel of fig.\ref{hdm} presents the observed DM relic abundance $\Omega_{X}h^{2}=0.12\pm 0.001$ \cite{Planck:2018vyg} projected to the plane of $m_{X}-\kappa$ by using the publicly available code \texttt{micrOMEGAs6.0} \cite{Alguero:2023zol}.
In this plot a bump appears near $m_{X}=m_{h/2}$, pointing to a change in the DM production process. 
Because in the DM mass range of $m_{X}<m_{h/2}$ the DM production is dominated by the decay $h\rightarrow \bar{X}X$ with $h$ the physical Higgs scalar,
but in the DM mass range of $m_{X}>m_{h/2}$, where the Higgs decay process is prohibited, 
the DM production mainly arises from the two-body annihilation of SM particles into $\bar{X}X$ via the virtual Higgs scalar. 
The required value of DM relic abundance helps us fix the value of $\kappa$, with $m_{X}$ being the only free variable.

The right panel of fig.\ref{hdm} shows the thermally averaged values of cross section of DM annihilation into $e\bar{e}$ as function of $m_X$, where the value of $\kappa$ is fixed as in the left panel of fig.\ref{hdm}. 
As the value of $m_{X}$ increases from $2m_{e}$ to $m_{h}/2$,  the magnitudes of $\left<\sigma v(X\bar{X}\rightarrow e\bar{e})\right>$ in units of cm$^{3}$/s range from $\sim 10^{-52}$ to $\sim 10^{-56}$, with a resonance taking place near $m_{X}=m_{h}/2$.
As the values of $\left<\sigma v(X\bar{X}\rightarrow e\bar{e})\right>$ are beyond the reach of both Lyman-$\alpha$ forest and 21-cm signal, 
this FIDM model will be no longer discussed in Sec.\ref{results}.

%%%%%%%%%%%%%%%%%%%%%%%%%%%%%%%%%%%%%%%%%%%%%%%%
%%%%%%%%%%%%%%%%%%%%%%%%%%%%%%%%%%%%%%%%%%%%%%%

\subsection{Neutrino portal}
\label{n}
The second FIDM model \cite{Asaka:2005cn,Becker:2018rve,Chianese:2018dsz,Datta:2021elq} is constructed through sterile neutrino with the following DM Lagrangian
\begin{eqnarray}\label{Nlag}
\mathcal{L}=\bar{N}_{I}(i\gamma^{\mu}\partial_{\mu}-m_{I})N_{I}-Y_{\alpha I} \bar{L}_{\alpha}\tilde{H}N_{I}
\end{eqnarray}
where $N_{I}$ with $I=$1-3  are the right-hand neutrinos ordered in mass, $L_{\alpha}=(\nu_{\alpha},\ell_{\alpha})^{T}$ with $\alpha=$1-3 the SM lepton doublets, and $\tilde{H}=i\sigma_{2}H^{*}$ with $H$ the Higgs doublet.  
In the situation where the lightest active neutrino mass is negligible,
the Yukawa coupling $Y_{\alpha 1}$ becomes small, allowing the DM candidate $N_1$ to freeze-in.

\begin{figure}
\centering
\includegraphics[width=8cm,height=8cm]{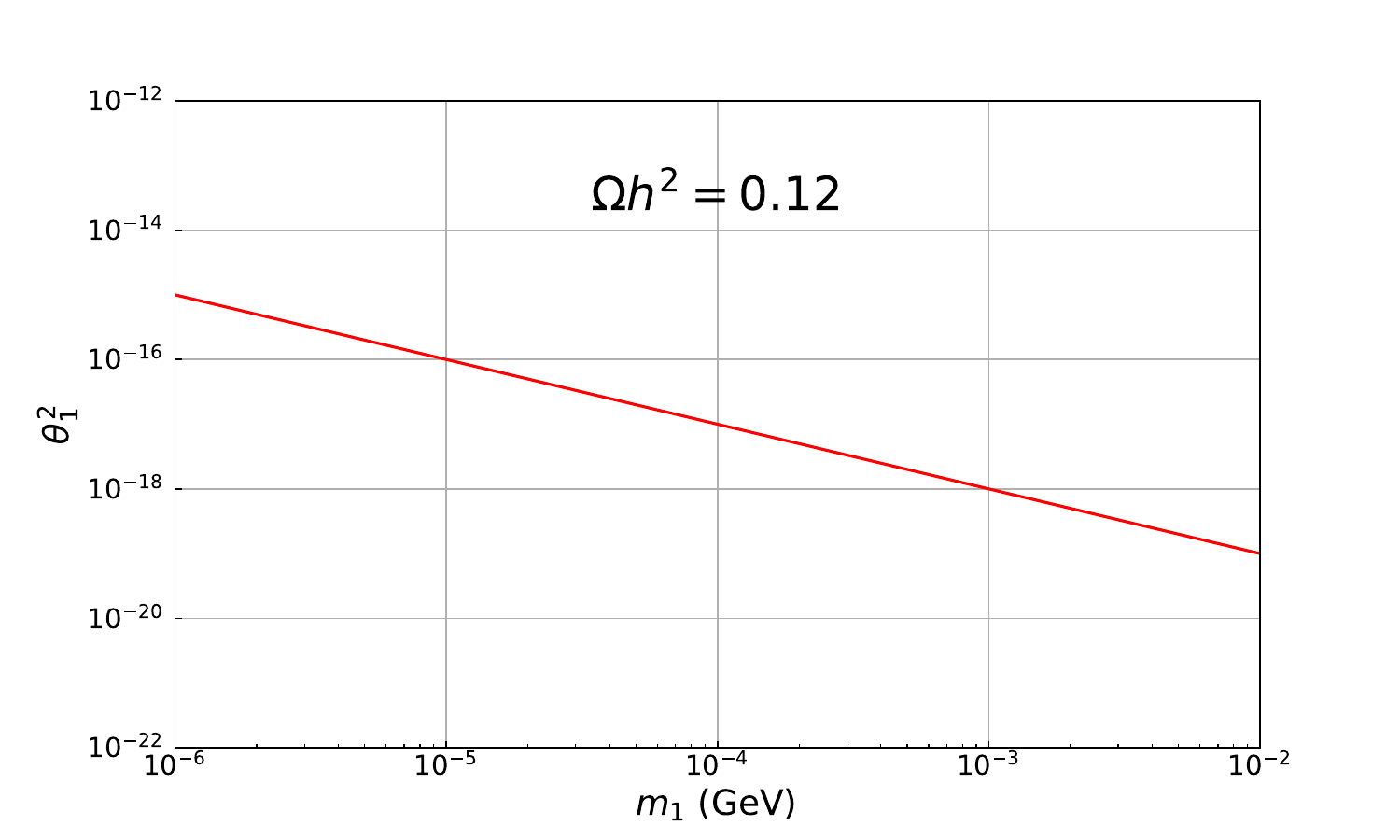}
\includegraphics[width=8cm,height=8cm]{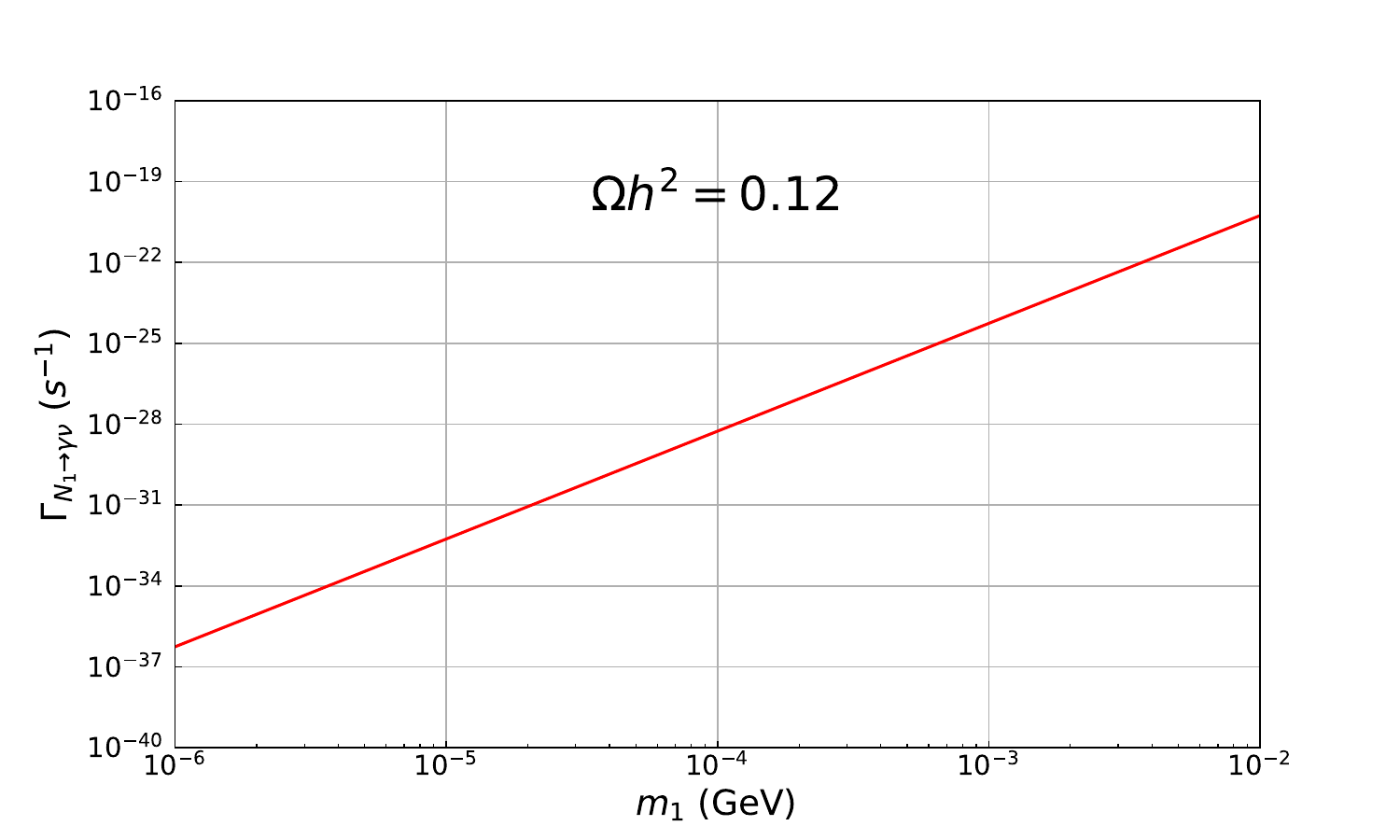}
\centering
\caption{$\mathbf{Left}$: the observed DM relic abundance $\Omega_{N_{1}}h^{2}=0.12\pm0.001$ projected to the plane of $m_{1}-\theta_{1}$. $\mathbf{Right}$: the decay width of $\Gamma_{N_{1}}$ as function of the DM mass $m_1$, 
where the value of $\theta_{1}$ has been fixed as in the left panel.}
\label{ndm}
\end{figure}

With the heavier sterile neutrinos $N_{2}$ and $N_3$ safely neglected, 
the freeze-in production of $N_1$ is dominated by $W^{\pm}\rightarrow N_{1}\ell^{\pm}_{\alpha}$.
The left plot of fig.\ref{ndm} shows the observed DM relic density projected to the plane of $m_{1}-\theta_{1}$ with 
$\theta^{2}_{1}=\sum_{\alpha=e, \mu, \tau}(Y^{2}_{\alpha1}\upsilon^{2}/2m^{2}_{1})$, which is consistent with the result of \cite{Datta:2021elq}. 
Using this plot to fix the value of $\theta_1$, we present the decay width of $N_1$ as function of the DM mass in the right plot of fig.\ref{ndm}, using \cite{Pal:1981rm,Barger:1995ty,Boyarsky:2009ix} 
\begin{eqnarray}\label{Nwidth}
\Gamma_{N_{1}}\approx \Gamma (N_{1}\rightarrow \gamma\nu)\approx \frac{9\alpha G^{2}_{F}}{1024\pi^{4}}\sin^{2}(2\theta_{1})m^{5}_{1}.
\end{eqnarray}
The right panel shows the magnitudes of $\Gamma_{N_{1}}$ in units of sec$^{-1}$ range from $\sim10^{-36}$ to $\sim 10^{-20}$ within the DM mass range of $m_{1}\sim (10^{-3}-10)$ MeV.
Note that this decay process transfers only a half of the DM rest mass energy into the IGM.

%%%%%%%%%%%%%%%%%%%%%%%%%%%%%%%%%%%%%%%%%%%%%%%%
%%%%%%%%%%%%%%%%%%%%%%%%%%%%%%%%%%%%%%%%%%%%%%%

%%%%%%%%%%%%% Millicharged DM %%%%%%%%%%%%%%%%%%%%%%%%%%
\subsection{Vector portal}
The third FIDM model is the so-called millicharged DM \cite{Holdom:1985ag,Feldman:2007wj,Yin:2023jql} with its Lagrangian given by
\begin{eqnarray}\label{DLag}
\mathcal{L}=i\bar{\psi}\gamma^{\mu}D_{\mu}\psi-m_{\psi}\bar{\psi}\psi,
\end{eqnarray}
where $D_{\mu}=\partial_{\mu}-i\epsilon g'B_{\mu}$ with $g'$ and $B_\mu$ the SM hypercharge gauge coupling and field, respectively,  
and $m_{\psi}$ is the DM mass.
Below the electroweak scale one obtains the interaction between the SM and DM sector 
\begin{eqnarray}\label{Dint}
\mathcal{L}_{\rm{int}}=\epsilon e \bar{\psi}\gamma^{\mu}\psi (A_{\mu}-\tan\theta_{W}Z_{\mu}).
\end{eqnarray}
where we have replaced $B_{\mu}=\cos\theta_{W}A_{\mu}-\sin\theta_{W}Z_{\mu}$ with $\theta_{W}$ the weak mixing angle.  
As mentioned in \cite{Yin:2023jql}, Eq.(\ref{Dint}) is more complete than the conventional millicharged DM studied in the literature, as
the coupling of DM to $Z$ has been also included, which affects the freeze-in production in the DM mass range of $m_{\psi}<m_{Z}/2$.

%%%%%%%%%%%%%%%%%%%%%%%%%%%%%%%%%%%%%%%%%%%%%%%%
\begin{figure}
\centering
\includegraphics[width=8cm,height=8cm]{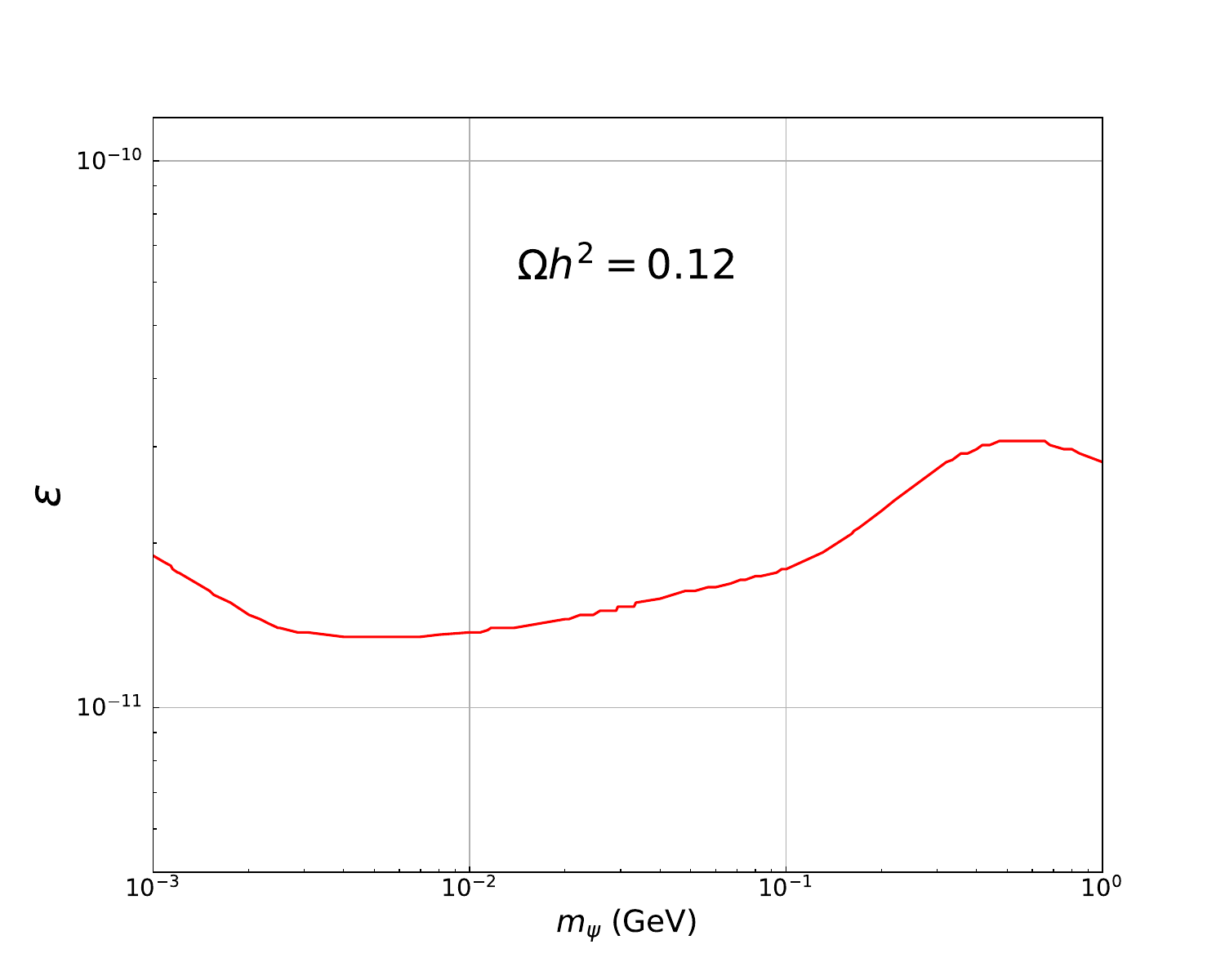}
\includegraphics[width=8cm,height=8cm]{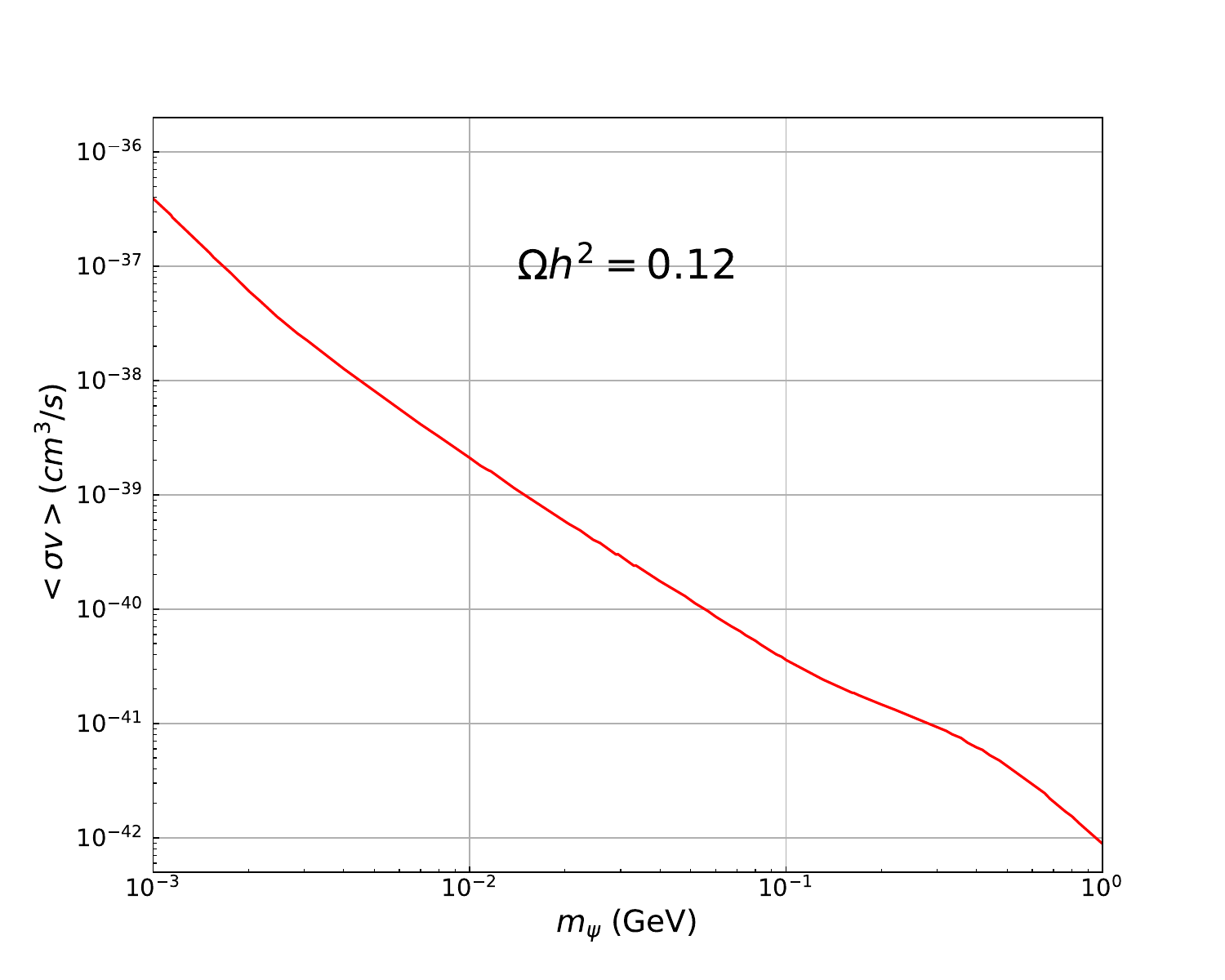}
\centering
\caption{$\mathbf{Left}$: the observed DM relic abundance $\Omega_{\psi}h^{2}=0.12\pm0.001$ projected to the plane of $m_{\psi}-\epsilon$. $\mathbf{Right}$:  the thermally averaged values of $\left<\sigma v(\psi\bar{\psi}\rightarrow e\bar{e})\right>$, 
with the value of $\epsilon$ fixed as in the left panel.}
\label{vdm}
\end{figure}
%%%%%%%%%%%%%%%%%%%%%%%%%%%%%%%%%%%%%%%%%%%%%%%

With a small $\epsilon$ the millicharged DM obtains its relic density via the freeze-in processes such as the decay $Z\rightarrow \psi\bar{\psi}$ 
and annihilation $f\bar{f}\rightarrow \psi\bar{\psi}$, depending on the value of $m_{\psi}$, where $f$ denotes the SM fermions. 
The left plot of fig.\ref{vdm} shows the observed DM relic density projected to the plane of $m_{\psi}-\epsilon$.
Compared to the millicharged DM without the $Z$ vertex in eq.(\ref{Dint}), 
the values of $\epsilon$ are slightly smaller with respect to a fixed value of $m_{\psi}$ being smaller than $m_{Z}/2$.
Eliminating $\epsilon$ with the plot of fig.\ref{vdm}, 
we present the values of $\left<\sigma v(\psi\bar{\psi}\rightarrow e\bar{e})\right>$  as function of $m_\psi$ in the right plot of fig.\ref{vdm},
showing that it rapidly decreases from $\sim 10^{-37}$ to $\sim 10^{-45}$  in units of  cm$^{3}/$s as $m_{\psi}$ increases from $\sim 1$ MeV to $\sim 1$ GeV. 
Therefore, the lower mass end of $m_{\psi}$ is more promising to leave footprints in the Lyman-$\alpha$ and 21-cm observations.

\section{Modeling dark matter induced effects on cosmological observations}
\label{modeling}
In this section we briefly discuss how to numerically calculate the values of Lyman-$\alpha$ and 21-cm observables from a viewpoint of phenomenology in Sec.\ref{Lyman} and Sec.\ref{21cmr} respectively,
where we will emphasize the effects of DM annihilation or decay induced energy injection into the IGM on these observables.

\subsection{Lyman-$\alpha$}
\label{Lyman}

In the late-time Universe the evolution of IGM ionization fraction and temperature is described by \cite{Liu:2019bbm} 
\begin{eqnarray}
\frac{dx_{\rm{HII}}}{dz}&=&\frac{dt}{dz}\left(\Lambda_{\rm{ion}}-\Lambda_{\rm{rec}}+\Lambda^{\rm{DM}}_{\rm{ion}}\right),\label{DE1}\\
\frac{dT_{k}}{dz}&=&\frac{2}{3}\frac{T_{k}}{n_{b}}\frac{dn_{b}}{dz}-\frac{T_{k}}{1+x_{e}}\frac{dx_{e}}{dz}+\frac{2}{3k_{B}(1+x_{e})}\frac{dt}{dz}\left(\sum_{p}\epsilon^{p}_{\rm{heat}}+\epsilon_{\rm{heat}}^{\rm{DM}}\right),\label{DE2}
\end{eqnarray}
where $x_{\rm{HII}}=n_{H^{+}}/n_{H}$ is the ionization fraction with $n_{H}$ ($n_{H^{+}}$) the number density of (ionized) hydrogen, $T_{k}$ the matter (baryon) temperature, $n_b$ the baryon number density, $k_B$ the Boltzmann constant, and $z$ the redshift.
In eq.(\ref{DE1}), $\Lambda_{\rm{ion}}$ includes the photoionization and astrophysical source induced ionization rate,
$\Lambda_{\rm{rec}}$ is the recombination rate, and $\Lambda^{\rm{DM}}_{\rm{ion}}$ represents the DM-induced ionization rate. 
In eq.(\ref{DE2}), $\epsilon^{p}_{\rm{heat}}$ includes the Compton scattering (effective at $z\geq 300$) and astrophysical source induced heating rate,
and $\epsilon_{\rm{heat}}^{\rm{DM}}$ is the DM induced heating rate. 

The DM annihilation or decay induced terms in eqs.(\ref{DE1}) and (\ref{DE2}) are given by \cite{Facchinetti:2023slb,Liu:2019bbm}
\begin{eqnarray}
\epsilon^{\rm{DM}}_{c}&=&f_{c}(x_{e},z)\frac{1}{n_{b}}\left(\frac{dE(z)}{dtdV}\right)_{\rm{inj}},\label{DMrate1}\\
\Lambda^{\rm{DM}}_{\rm{ion}}&=& \mathcal{F}_{\rm{H}}\frac{\epsilon^{\rm{DM}}_{\rm{HII}}}{E^{\rm{HI}}_{\rm{th}}}+\mathcal{F}_{\rm{He}}\frac{\epsilon^{\rm{DM}}_{\rm{HeII}}}{E^{\rm{HeI}}_{\rm{th}}},\label{DMrate2}
\end{eqnarray}
where $f_{c}(x_{e},z)$ are the deposition fractions, with deposition channels including IGM heating (c=heat), hydrogen ionization (c = HII), helium single or double ionization (c = HeII or HeIII), and neutral atom excitation (c = exc), 
$\mathcal{F}_{j}$ refers to the number fraction of each species $j$, $E^{j}_{\rm{th}}$ is the energy for ionization.
In eq.(\ref{DMrate1}) the DM induced energy injection rate is defined as
\begin{eqnarray}\label{DMrate}
\left(\frac{dE(z)}{dtdV}\right)_{\rm{inj}}=
\left\{
\begin{array}{lcl}
\rho^{2}_{\rm{DM},0}(1+z)^{6}\left<\sigma v\right>/m_{\rm{DM}}, ~~~~~~~\rm{annihilation}\\
\rho_{\rm{DM},0}(1+z)^{3}\Gamma_{\rm{DM}}, ~~~~~~~~~~~~~~~~\rm{decay}\\
\end{array}\right.
\label{Lint}
\end{eqnarray}
where $\left<\sigma v\right>$ is the velocity-averaged DM annihilation cross section, $\Gamma_{\rm{DM}}$ the DM decay width, 
$\rho_{\rm{DM},0}$ the present value of DM relic density, and $m_{\rm{DM}}$ the DM mass.

Given an explicit DM model, the energy injection rates in eq.(\ref{DMrate}) are specified as illustrated in Sec.\ref{models}.
Taking these rates as inputs, we use the publicly available package \texttt{DarkHistory} \cite{Liu:2019bbm,Sun:2022djj} to calculate the deposition fractions $f_{c}(x_{e},z)$ in eq.(\ref{DMrate1}) and to derive the Lyman-$\alpha$ limits on the DM annihilation or decay rate. 
In particular, \texttt{DarkHistory} 
\begin{itemize}
\item uses $x_{\rm{e}}=n_{\rm{e}}/n_{H}$, with $n_{\rm{e}}$ the number density of  free electron;
\item chooses the case-B photoionization and recombination coefficients for hydrogen;
\item allows us to parametrize the astrophysical source contributing to photoionization and photoheating. 
\end{itemize}

\subsection{21-cm signal}
\label{21cmr}
Now we turn to 21-cm cosmology.\footnote{For a review see \cite{Pritchard:2011xb}.}
Instead of eqs.(\ref{DE1}) and (\ref{DE2}), it is more common to use 
\begin{eqnarray}
\frac{dx_{e}}{dz}&=&\frac{dx^{\rm{DM}}_{e}}{dz}+\frac{dt}{dz}\left(\Lambda_{\rm{ion}}-\alpha_{A}Cx^{2}_{e}n_{A}f_{H}\right), \label{DE3}\\
\frac{dT_{k}}{dz}&=&\frac{dT_{k}^{\rm{DM}}}{dz}+\frac{2}{3k_{B}(1+x_{e})}\frac{dt}{dz}\sum\epsilon^{p}_{\rm{heat}}+\frac{2T_{k}}{3n_{A}}\frac{dn_{A}}{dz}-\frac{T_{k}}{1+x_{e}}\frac{dx_{e}}{dz},\label{DE4}
\end{eqnarray}
where $\alpha_{A}$ is the case-A recombination coefficient, 
$C$ is the free-electron clumping factor, $n_A$ is the local local physical nuclear density, 
$f_{H}$ is the hydrogen nucleus number fraction,
and $dx^{\rm{DM}}/dz$ and $dT^{\rm{DM}}_{k}/dz$ represent the DM induced effects similar to $\Lambda^{\rm{DM}}_{\rm{ion}}$ in eq.(\ref{DE1}) and $\epsilon^{\rm{DM}}_{\rm{heat}}$ in eq.(\ref{DE2}) respectively.

Moreover, the Wouthuysen-Field coupling $x_{\alpha}$ is also modified by any exotic energy injection involved,
\begin{eqnarray}\label{Lymanc}
x_{\alpha}=1.7\times10^{11}\frac{S_{\alpha}}{1+z}\frac{J_{\alpha}}{\rm{s}^{-1}\rm{Hz}^{-1}\rm{cm}^{-2}\rm{sr}^{-1}}
\end{eqnarray}
where $S_{\alpha}$ is a correction coefficient \cite{Hirata:2005mz} of order unity, 
and $J_{\alpha}$ is the Lyman-$\alpha$ background density
\begin{eqnarray}\label{J}
J_{\alpha}\rightarrow J_{\alpha}+J_{\alpha}^{\rm{DM}},
\end{eqnarray}
with DM induced Lyman-$\alpha$ intensity \cite{Facchinetti:2023slb}
\begin{eqnarray}\label{JDM}
J_{\alpha}^{\rm{DM}}=\frac{cn_{b}}{4\pi H(z)\nu_{\alpha}}\frac{\epsilon_{\rm{Ly}\alpha}}{h\nu_{\alpha}}.
\end{eqnarray} 
Here, $H$ is the Hubble rate, $\nu_{\alpha}$ the Lyman-$\alpha$ frequency and $\epsilon_{\rm{Ly}\alpha}$ the DM induced Lyman-$\alpha$ excitation.

Using eqs.(\ref{DE3}), (\ref{DE4}) and (\ref{Lymanc}), one is able to derive the effects of DM induced energy injection into the IGM on the spin temperature 
\begin{eqnarray}\label{TS}
T^{-1}_{S}=\frac{T^{-1}_{\rm{CMB}}+(x_{\alpha}+x_{c})T^{-1}_{k}}{1+x_{\alpha}+x_{c}},
\end{eqnarray}
where $x_{c}$ is the collision coupling.
Given the value of $T_S$ in eq.(\ref{TS}), it is straightforward to determine the differential brightness temperature of 21-cm signal arising from the hyperfine spin-flip transition of neutral hydrogen \cite{Furlanetto:2006jb}
\begin{eqnarray}\label{Tb}
\delta T_{21}(z)\approx 20x_{\rm{HI}}(1+\delta_{b})\left(1-\frac{T_{\rm{CMB}}}{T_{S}}\right) \left(\frac{1+z}{10}\right)^{1/2}\left(\frac{\Omega_{b}h^{2}}{0.023}\right)\left(\frac{\Omega_{m}h^{2}}{0.15}\right)^{-1/2}\rm{mK},
\end{eqnarray}
where $x_{\rm{HI}}$ is the neutral hydrogen fraction, $\delta_b$ is the baryon over density, $\Omega_{b}$ and $\Omega_m$ are the baryon and matter energy density relative to the present critical density respectively, and the Hubble parameter $h$ is defined as $H_{0}=h\cdot 100$ km s$^{-1}$Mpc$^{-1}$ with $H_0$ the present-day Hubble rate.

Apart from the global 21-cm signal, spatial variation of IGM quantities leads to fluctuations in the 21-cm signal.
The 21-cm power spectrum is defined as
\begin{eqnarray}\label{21ps}
\overline{\delta T^{2}_{21}}\Delta^{2}_{21}(k,z)=\overline{\delta T^{2}_{21}(z)}\times \frac{k^{3}}{2\pi^{2}}P_{21}(k,z)
\end{eqnarray}
where $\overline{\delta T_{21}}$ is the sky-averaged brightness temperature and $P_{21}$ is given by
\begin{eqnarray}\label{ps}
\left<\tilde{\delta}_{21}(\mathbf{k},z)\tilde{\delta}_{21}(\mathbf{k}',z)\right>=(2\pi)^{3}\delta^{D}(\mathbf{k}'-\mathbf{k})P_{21}(k,z)
\end{eqnarray}
with $\tilde{\delta}_{21}(\mathbf{k},z)$ the Fourier transformation of $\delta_{21}(\mathbf{x},z)=\delta T_{21}(\mathbf{x},z)/\overline{\delta T_{21}}(z)-1$.

To derive 21-cm limit on DM induced energy injection, we use the package \texttt{DM21cm} \cite{Sun:2023acy}\footnote{The current version of this package is only viable to deal with DM decay induced energy injection into the IGM.}  which combines \texttt{DarkHistory} and \texttt{21cmFAST} \cite{Mesinger:2010ne,Murray:2020trn}.
Explicitly, \texttt{DM21cm}
\begin{itemize}
\item follows the convention $x_{e}=n_{e}/n_{b}$;
\item chooses the case-A photoionization and recombination coefficients for hydrogen;
\item parametrizes the astrophysical source in terms of modeling star formation.
\end{itemize}
\texttt{DM21cm} uses \texttt{DarkHistory} to calculate transfer functions which depend on $\delta_{b}$, $x_{\rm{HI}}$, $T_k$, incident photon flux etc for each location in the simulation volume at redshift $z_i$. Over the interval $z_{i}$ to $z_{i+1}$, 
the transfer functions are then used to generate a new uniform photon bath and X-ray luminosity field,
and the energy deposition fields obtained from \texttt{DarkHistory} are combined with \texttt{21cmFAST} to yield a new simulation state of \texttt{21cmFAST} containing the information of $x_e$, $T_k$ and $\delta_{b}$.  
Repeating this process, we obtain the evolution of those IGM parameters as function of $z$.
With respect to each $z$,
the 21-cm power spectrum $\Delta^{2}_{21}(k,z)$ is derived in terms of \texttt{21cmFAST}.

%%%%%%%%%%%%%%%%%%%%%%%%%%%%
\section{Results}
\label{results}
As mentioned in Sec.\ref{h}, the Higgs-portal FIDM model is beyond the reach of both Lyman-$\alpha$ forest and 21-cm signal.
In this section, we only present these limits on the neutrino- and vector-portal FIDM model.

\subsection{Neutrino portal}
\subsubsection{Lyman-$\alpha$}
\label{nLy}
\begin{figure}
\centering
\includegraphics[width=8cm,height=8cm]{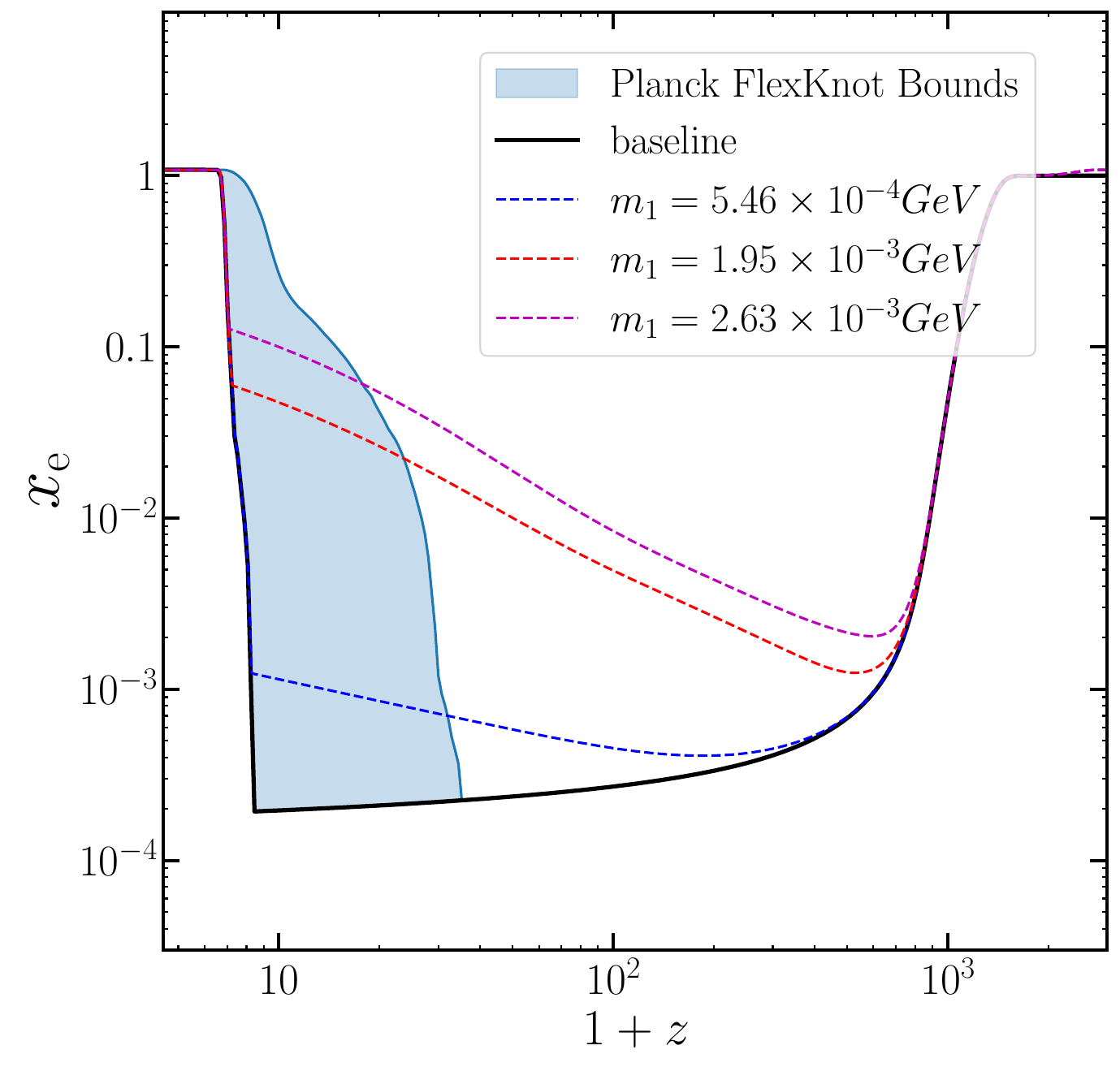}
\includegraphics[width=8cm,height=8cm]{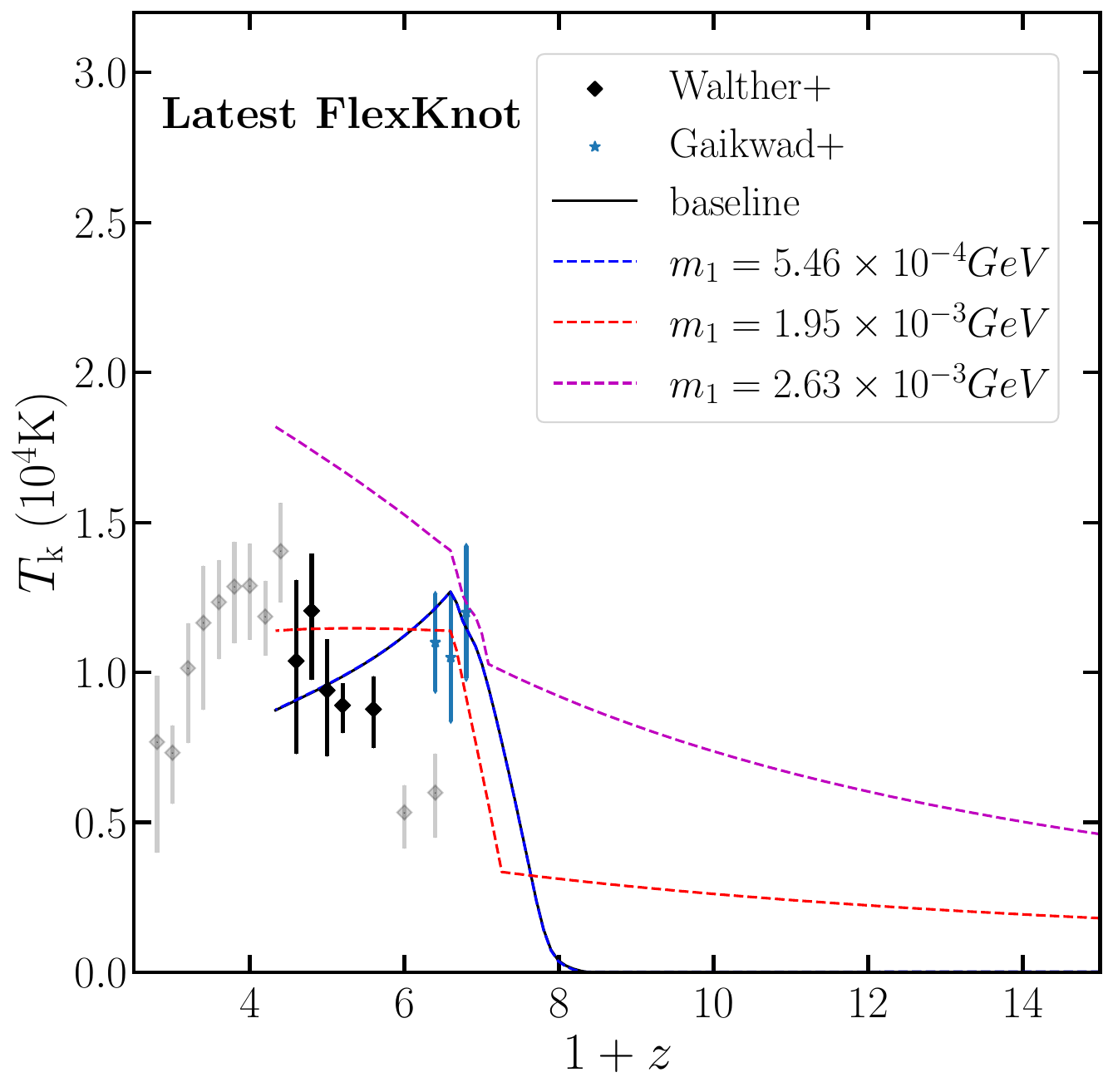}\\
\centering
\caption{The baseline ionization history (in black) of $x_e$ (left) and $T_k$ (right) based upon the FlexKnot model and photoheated parametrization.
Besides the baseline ionization history, we also present the sterile neutrino DM decay induced effects on the evolution of these two IGM parameters for various values of DM mass $m_{1}=\{0.546, 1.06, 1.95, 2.40, 2.63\}$ MeV using fig.\ref{ndm}.
While being consistent with current JWST data \cite{Umeda:2023} on $x_e$ in the left plot,  
these deviations have been constrained by the data \cite{Walther:2018pnn,Gaikwad:2020art} on $T_k$ as seen in the right plot. See text for details.}
\label{nLyman}
\end{figure}

Fig.\ref{nLyman} shows the baseline ionization history (in black) of  $x_{e}$ (left) and $T_{k}$ (right) as function of redshift $z$.  
The baseline ionization history is inferred from the astrophysical reionization constrained by the Planck data \cite{Planck:2018vyg} about $x_e$ and the data  \cite{Walther:2018pnn,Gaikwad:2020art} about $T_{k}$ as follows.
Similar to \cite{Liu:2020wqz} we adopt the FlexKnot model to parametrize the astrophysical source contribution to $\Lambda_{\rm{ion}}$ in eq.(\ref{DE1}) and a photoheated prescription to parametrize the astrophysical source contribution to $\epsilon^{p}_{\rm{heat}}$ in eq.(\ref{DE2}) simultaneously. 
Using the data \cite{Walther:2018pnn,Gaikwad:2020art} on $T_{k}$ in the low redshift range of $z\sim 4-7$,
the parameters $(\Delta T, \alpha_{\rm{bk}})$ in the photoheated prescription can be fixed \cite{Liu:2020wqz}. 
So, a combination of the Planck data \cite{Planck:2018vyg} and the data \cite{Walther:2018pnn,Gaikwad:2020art} enables us to determine the astrophysical reionization. 

Alongside the baseline ionization history, in fig.\ref{nLyman} we also show the evolution of $x_{e}$ and $T_{k}$ with the neutrino-portal FIDM decay induced energy injection into the IGM taken into account for various values of DM mass $m_{1}=\{0.546,1.95, 2.63\}$ MeV using the right plot of fig.\ref{ndm}.
Despite being consistent with current JWST data \cite{Umeda:2023},
these deviations have been constrained as seen in the plot of $T_k$, 
since compared to the astrophysical contribution the DM decay induced contribution to $T_k$ should be smaller.

Compared to the Lyman-$\alpha$ limit in \cite{Liu:2020wqz} where the astrophysical contribution to reionization changes depending on the DM induced contribution in order to fit the Walther$+$ and Gaikwad$+$ data,
here the astrophysical contribution has been fixed by the Walther$+$ and Gaikwad$+$ data together with the Planck data.
Since in the context of astrophysical reionization considered, 
the DM induced contribution to reionization should be smaller compared to the astrophysical contribution. 
Therefore, the Lyman-$\alpha$ limits derived below are not a recasting of those of \cite{Liu:2020wqz} but expected to be stronger.

%%%%%%%%%%%%%%%%%%%%%%%%%%%%%%%%%%%%%%%%%%%%%%%%
%%%%%%%%%%%%%%%%%%%%%%%%%%%%%%%%%%%%%%%%%%%%%%%

%%%%%%%%%%%%%%%%%%%%%%%%%%%%%%%%%%%%%%%%%%%%%%%%
\begin{figure}[htb!]
\centering
\includegraphics[width=8cm,height=8cm]{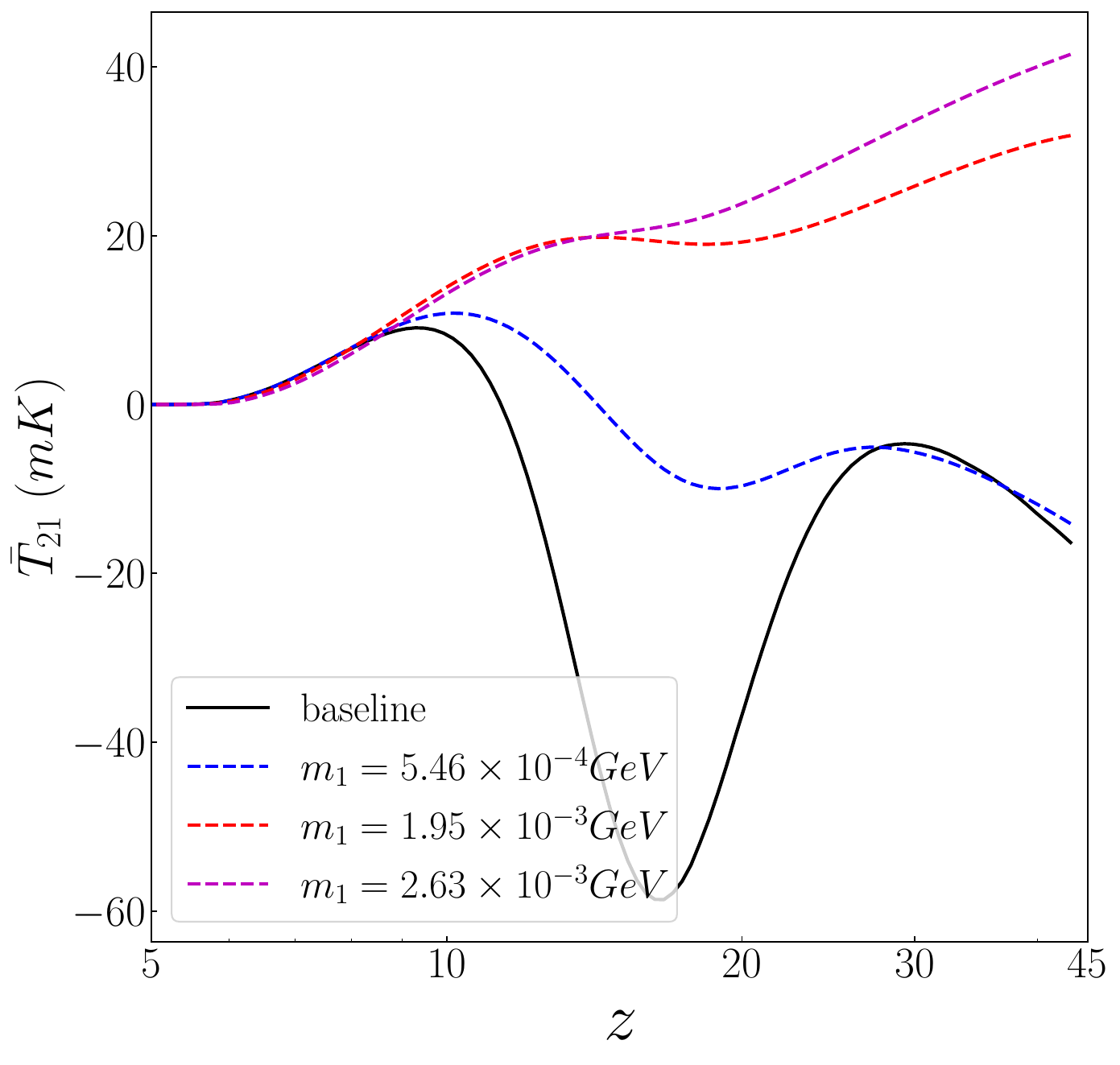}
\includegraphics[width=8cm,height=8cm]{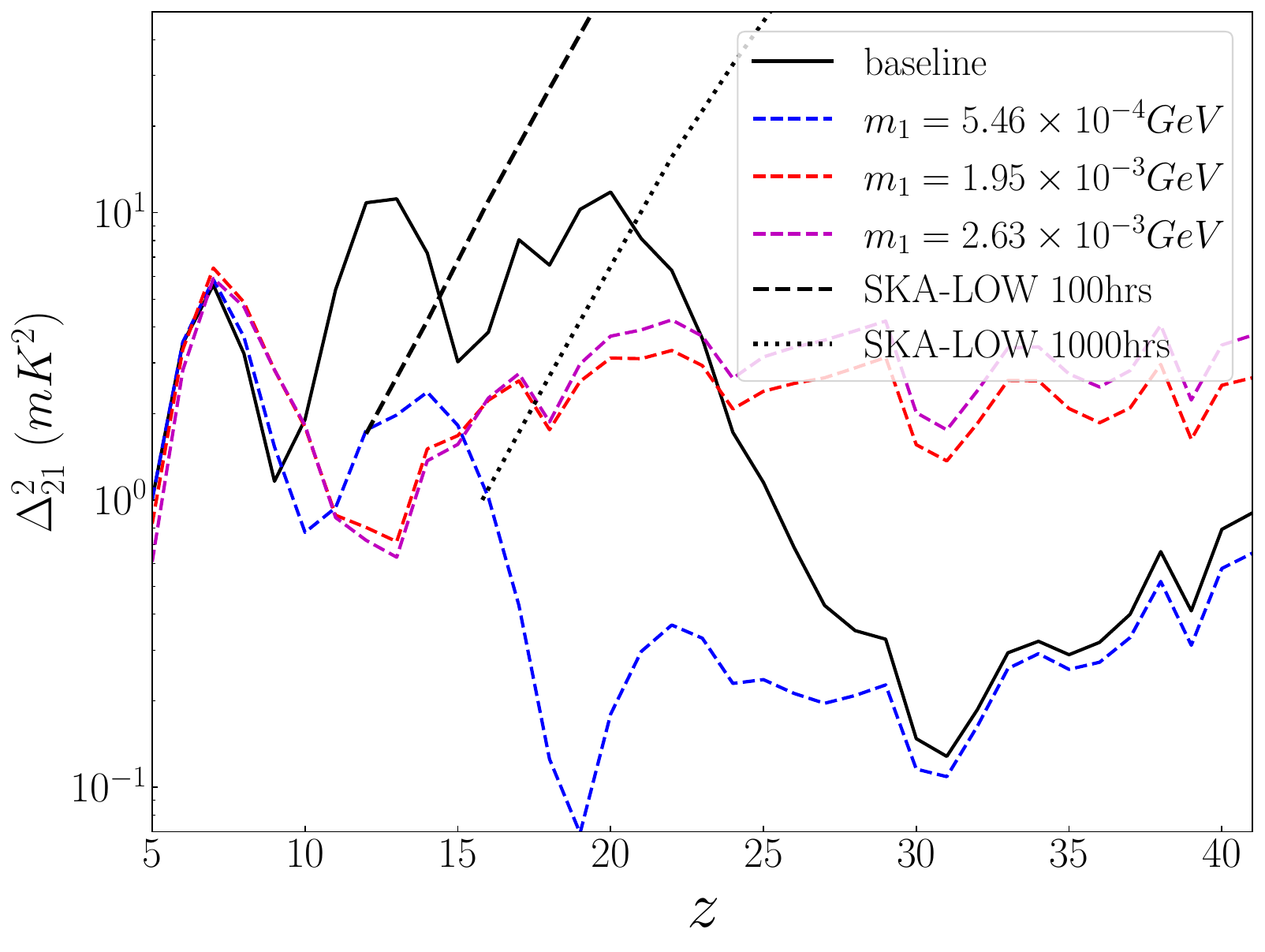}
\centering
\caption{$\mathbf{Left}$: the baseline values of $\left<T_{21}\right>$ (in black) as function of $z$ and the deviations from them due to the sterile neutrino DM decay induced effects on the IGM for various values of DM mass $m_{1}$. 
$\mathbf{Right}$: the same as the left panel but for the 21-cm power spectrum $\Delta^{2}_{21}$ with respect to the reference wavenumber $k_{*}=0.2h$ Mpc$^{-1}$, compared to future SKA \cite{Mertens:2021apf} sensitivities with respect to $100$ hrs (in black dashed) and $1000$ hrs (in black dotted)  simultaneously, which implies that a precision of order $\sim 10\%$ in $\Delta^{2}_{21}(k_{*})$ in the redshift range of $z\sim 15-16$ provided by SKA ($1000$ hrs) is sufficient to exclude the DM mass range of $m_{1}\geq 0.546$ MeV.}
\label{21cm}
\end{figure}
%%%%%%%%%%%%%%%%%%%%%%%%%%%%%%%%%%%%%%%%%%%%%%%

\subsubsection{21-cm signal}
The $\mathbf{left}$ panel of fig.\ref{21cm} shows the global values of $\left<T_{21}\right>$ as function of $z$. 
In this plot the baseline values of $\left<T_{21}\right>$ (in black) arises from the standard astrophysical processes such as stellar emission of UV and X-ray photons leading to energy deposition into heating, ionization and Lyman-$\alpha$ excitation, which are modeled by \texttt{21cmFAST} with fiducial values of \texttt{21cmFAST} parameters taken from \cite{Munoz:2021psm,Mason:2022obt}.
Meanwhile, this plot also shows the FIDM decay induced deviations from the baseline values of $\left<T_{21}\right>$ with respect to various values of DM mass using fig.\ref{ndm}.
While EDGES \cite{Bowman:2018yin} reported a measurement on $\left<T_{21}\right>$, it however disputes with SARAS3 \cite{Singh:2022ivh} among others.
Therefore, we do not make use of this data for the present analysis but instead consider the sensitivities of future experiments on the 21-cm power spectrum as below.

The $\mathbf{right}$ panel of fig.\ref{21cm}  shows the values of the 21-cm power spectrum $\Delta^{2}_{21}$ as function of $z$ with respect to the reference wavenumber $k_{*}=0.2h$ Mpc$^{-1}$. 
Explicitly, we have chosen a comoving volume of $(256 \ \text{Mpc})^3$ with a comoving grid resolution of 2 Mpc and a redshift interval of $\delta z=1$
for the $T_{21}$ lightcones.
This panel shows that the FIDM decay induced deviations from the baseline values of $\Delta^{2}_{21}(k_{*})$ are small in low redshift region ($z\leq 10$) but large in the high redshift range of $z>10$.
This feature is consistent with the previous results of \cite{Sun:2023acy}.
Consider that these deviations are far below current LOFAR \cite{Patil:2017zqk,Mertens:2020llj} and HERA  \cite{HERA:2021bsv,HERA:2022wmy} limits, 
we compare them to future HERA \cite{Munoz:2018jwq} and SKA \cite{Mertens:2021apf,Koopmans:2015sua} experiment. 
While beyond the expected sensitivity of HERA  which is not shown here,
the DM mass range of $m_{1}\geq 0.546$ MeV  can be excluded by a precision of order $\sim 10\%$ in $\Delta^{2}_{21}(k_{*})$ in the redshift range of $z\sim 15-16$ provided by SKA-low ($1000$ hrs).

 %%%%%%%%%%%%%%%%%%%%%%%%%%%%%%%%%%%%%%%%%%%%%%%%
\begin{figure}
\centering
\includegraphics[width=16cm,height=10cm]{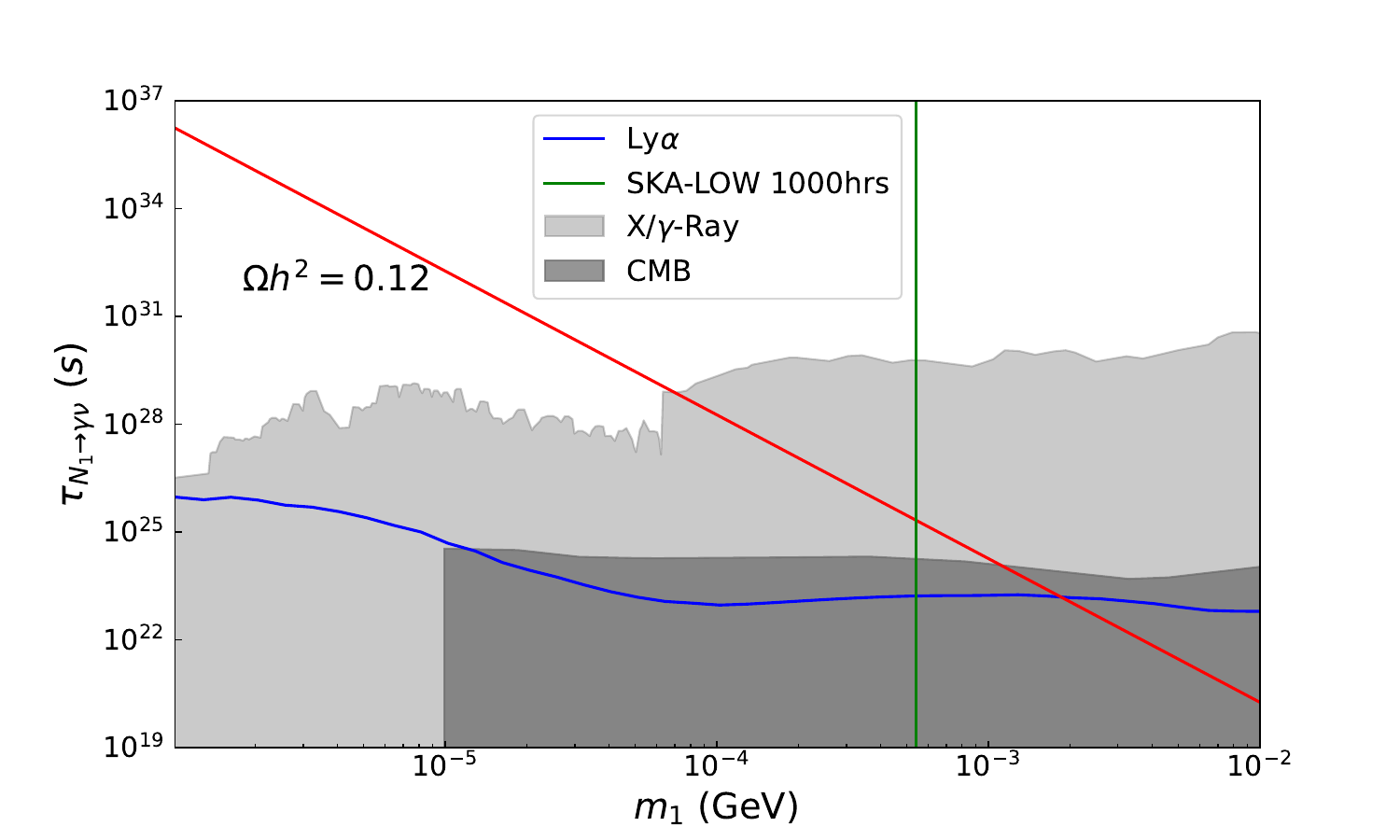}
\centering
\caption{The Lyman-$\alpha$ limit at 95$\%$ CL and the forecast sensitivity of SKA-low (1000 hrs) on the parameter space of sterile neutrino FIDM,
compared to other existing bounds including CMB \cite{Slatyer:2016qyl} and $X/\gamma$-ray \cite{Essig:2013goa, Horiuchi:2013noa, Ng:2019gch, Roach:2022lgo, Cirelli:2023tnx} at 95$\%$ CL.}
\label{nfinal}
\end{figure}
%%%%%%%%%%%%%%%%%%%%%%%%%%%%%%%%%%%%%%%%%%%%%%%

Compared to the 21-cm limits in \cite{Sun:2023acy} which are obtained in terms of HERA sensitivities on the 21-cm power spectrum within the range of $k\sim 0.1-1$ Mpc$^{-1}$, with the values of \texttt{21cmFAST} parameters allowed to vary,
we have instead fixed the values of \texttt{21cmFAST} parameters and replaced the HERA sensitivities by the SKA sensitivities.
Therefore, our 21-cm limit is stronger than those in \cite{Sun:2023acy} due to improved sensitivities provided by the SKA experiment.

\subsubsection{Comparison with existing limits}
Fig.\ref{nfinal} shows the Lyman-$\alpha$ limit at 95$\%$ CL (in blue)  and forecast sensitivity of SKA-low (1000 hrs) (in green) on the parameter space of sterile neutrino DM,
compared to the existing bounds on DM decay into photons from CMB \cite{Slatyer:2016qyl} and $X/\gamma$-ray telescopes \cite{Essig:2013goa, Horiuchi:2013noa, Ng:2019gch, Roach:2022lgo, Cirelli:2023tnx} at 95$\%$ CL.
In this figure, the Lyman-$\alpha$ limit has been derived by using  $\tau_{N_{1}}$ and $m_{1}$ as free input parameters 
which  is thus model-independent;
the 21-cm limit has used the result of fig.\ref{21cm} which is only valid for the sterile neutrino FIDM;
and all of other existing limits have been properly recasted.
This figure shows that the Lyman-$\alpha$, CMB, SKA-low (1000 hrs) and $X/\gamma$-ray, 
ordered in the ability of exclusion, excludes the DM mass above $1.8\times 10^{-3}$ GeV, $10^{-3}$ GeV,  $5.46\times 10^{-4}$ GeV and $6\times 10^{-5}$ GeV, respectively.
To summarize, the DM mass with $m_{1}\leq 60$ keV still survives.

%%%%%%%%%%%%%%%%%%%%%%%%%%%%%%%%%%%%%%%%%%%%%%%%
\begin{figure}
\centering
\includegraphics[width=8cm,height=8cm]{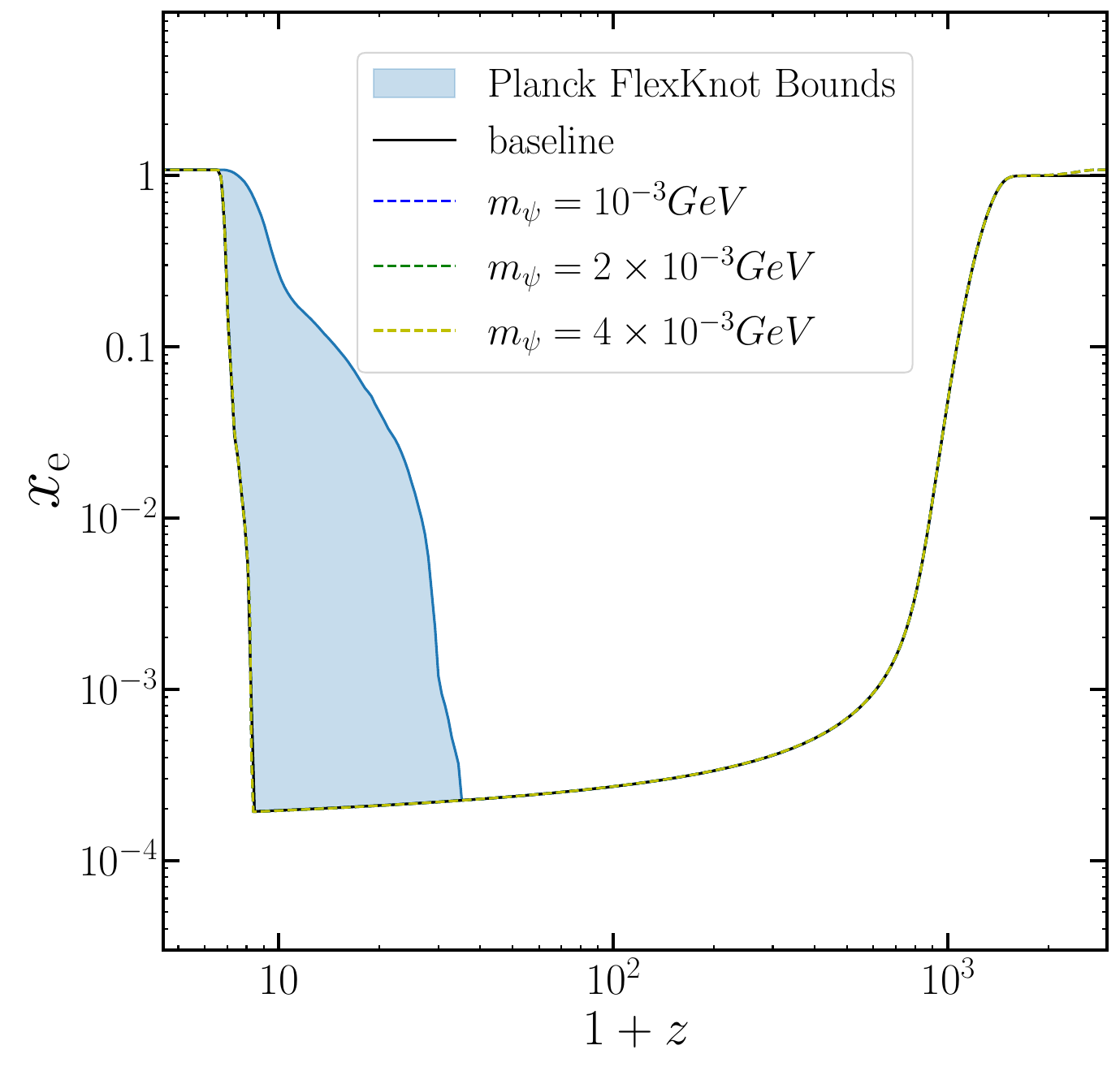}
\includegraphics[width=8cm,height=8cm]{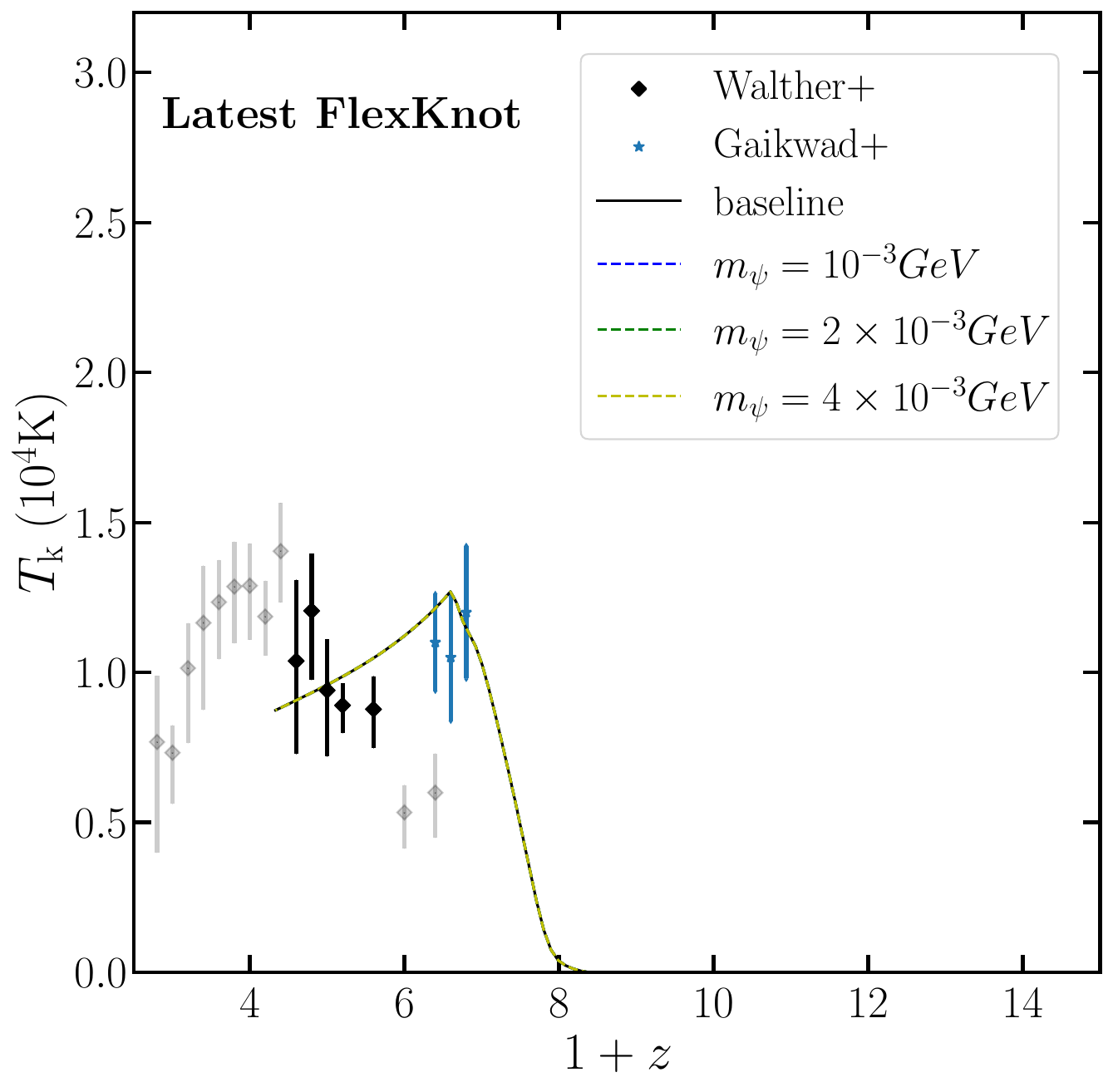}\\
\centering
\caption{Same as fig.\ref{nLyman} but for the millicharged FIDM, where 
we show this DM annihilation induced effects on the evolution of the IGM parameters for various values of DM mass $m_{\psi}=\{1,2,4\}$ MeV using fig.\ref{vdm}.}
\label{vLyman}
\end{figure}
%%%%%%%%%%%%%%%%%%%%%%%%%%%%%%%%%%%%%%%%%%%%%%%

%%%%%%%%%%%%% dark photon %%%%%%%%%%%%%%%%%%%%%%%%%%
\subsection{Vector portal}
\subsubsection{Lyman-$\alpha$}

Fig.\ref{vLyman} shows the baseline ionization history (in black) of $x_e$ (left) and $T_k$ (right) based upon the FlexKnot model and photoheated parametrization, and the millicharged FIDM annihilation induced effects on the evolution of these IGM parameters for various values of DM mass $m_{\psi}=\{1,2,3\}$ MeV using fig.\ref{vdm}.
In this figure, the deviations relative to the base line values are of order $\leq 10^{-5}$ and $\leq 10^{-4}$ for $x_e$ and $T_k$, respectively, as seen in fig.\ref{vdm} where the values of $\left<\sigma v(\psi\bar{\psi}\rightarrow e\bar{e})\right>$ in the DM range of $10^{-3}-1$ GeV are at least several orders of magnitude smaller than $\sim 10^{-30}\rm{cm}^{3}/$s - the threshold value reached by the current Lyman-$\alpha$ data.
This implies that the Lyman-$\alpha$ limit presented below is hardly to place a constraint on the millicharged FIDM.
Nevertheless, this new model-independent limit is still useful, 
because it excludes a large portion of the millicharge parameter space composed of $m_{\psi}$ and $\epsilon$ as a complementary probe.

\subsubsection{Comparison with existing limits}

\begin{figure}
\centering
\includegraphics[width=15cm,height=10cm]{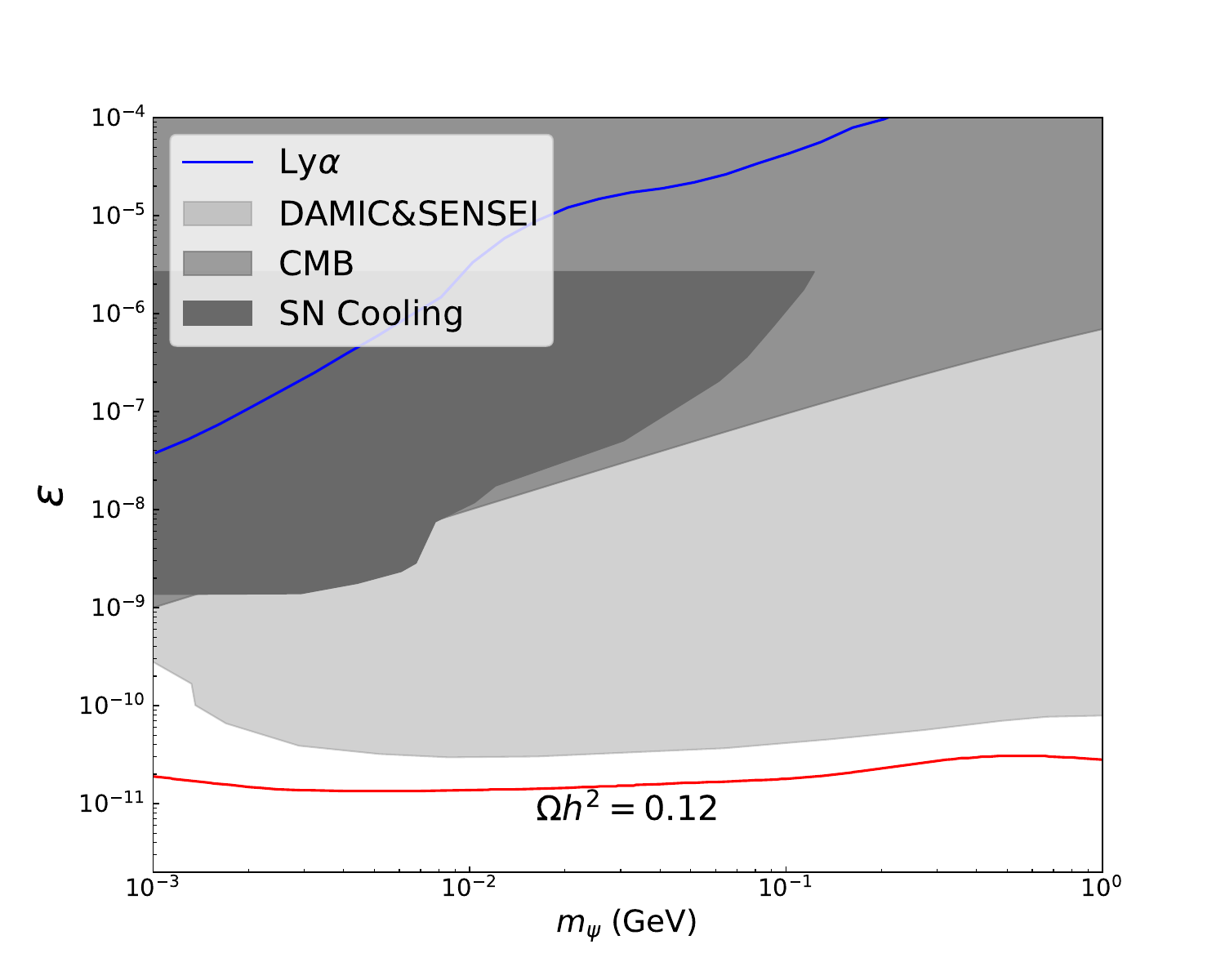}
\centering
\caption{The Lyman-$\alpha$ limit (in blue) on the millicharge parameter space at 95$\%$ CL, 
compared to the existing bounds from CMB \cite{Slatyer:2016qyl}, SN 1987A \cite{Chang:2018rso} and DM direct detections \cite{Iles:2024zka} from SENSEI \cite{SENSEI:2023zdf} and DAMIC \cite{DAMIC-M:2023gxo}.}
\label{vfinal}
\end{figure}

Finally we show in fig.\ref{vfinal} the Lyman-$\alpha$ limit (in blue) on the millicharged parameter space at 95$\%$ CL,
compared to the existing bounds including CMB \cite{Slatyer:2016qyl}, Supernova (SN) 1987A \cite{Chang:2018rso} 
and DM direct detections \cite{Iles:2024zka} on DM-electron scattering from SENSEI \cite{SENSEI:2023zdf} and DAMIC \cite{DAMIC-M:2023gxo}.
As mentioned above, the Lyman-$\alpha$ limit in this plot is model-independent,
which only excludes the millicharge $\epsilon$ down to $\sim 10^{-8}$ within the DM mass range of $10^{-3}-1$ GeV.
To summarize, the surviving parameter space (in red) of millicharged FIDM is still intact,
which can be reached by the near future DM direct detections.

 %%%%%%%%%%%%%%%%%%%%%%%%%%%%%%%%%%%%%%%%%%%%%%%%
%%%%%%%%%%%%%%%%%%%%%%%%%%%%%%%%%%%%%%%%%%%%%%%

%%%%%%%%%%%%%%%%%%%%%%%%%%%%
\section{Conclusion}
\label{con}

In this work we have reported new model-independent Lyman-$\alpha$ and model-dependent 21-cm constraints on two single-field FIDM models.
Regarding the Lyman-$\alpha$, we have fixed the baseline ionization history using low redshift data about astrophysical reionization,
whereas for 21-cm signal we have adopted the baseline values of 21-cm power spectrum through a standard modeling of star formation developed so far.
Using the latest numerical tools, we have shown that (i) for sterile neutrino FIDM, current Lyman-$\alpha$ data and future sensitivity of SKA-low (1000 hrs) on the 21-cm power spectrum with respect to $k_{*}=0.2h$ Mpc$^{-1}$ at the redshift range of $z\sim 15-16$ excludes the DM mass up to $1.8\times 10^{-3}$ GeV at 95$\%$ CL and $5.46\times 10^{-4}$ GeV, respectively, 
and (ii) for millicharged FIDM, current Lyman-$\alpha$ data has only excluded the millicharge $\epsilon$ down to $\sim 10^{-8}$ within the DM mass range of $10^{-3}-1$ GeV at 95$\%$ CL, 
which implies that the surviving parameter space of millicharged FIDM is still intact.

Several factors affect the derived exclusions.
For the Lyman-$\alpha$ constraints, 
the baseline ionization history with respect to the astrophysical reionization
relies on the Walther$+$ and Gaikwad$+$ data. 
Using a set of data points different from \cite{Liu:2020wqz} which we have followed here may mildly change the baseline ionization history.
Alternatively, one can even replace the astrophysical reionization by DM reionization.
In this situation,
the Walther$+$ and Gaikwad$+$ data allow a larger DM contribution to $T_k$ at $z\sim 4-6$,
which implies a larger $x_e$ at high redshift region accordingly. 
For the 21-cm constraints, 
\texttt{DM21cm} has followed \texttt{21cmFAST} to adopt two stellar populations, each of which contains several parameters. 
If one takes fiducial values of the \texttt{21cmFAST} parameters different from \cite{Munoz:2021psm,Mason:2022obt},
the baseline values of 21-cm power spectrum are expected to be modified.
Finally,
the annihilation or decay of FIDM into SM final states rather than $e\bar{e}$ and photons may indirectly contribute to the FIDM induced deviations from the baseline values of Lyman-$\alpha$ and 21-cm observables.

Our approach can be applied to other FIDM models. 
Take axion-like DM for example.
It mainly decays into photons similar to the neutrino-portal FIDM considered here.
Both the Lyman-$\alpha$ and 21-cm constraints on the axion-like DM can be similarly derived.
Likewise, this approach can be also applied to two-field FIDM models where DM annihilates or decays into $e\bar{e}$ or photons.

\section*{Data availability statement}
Codes used in this work can be found in \cite{Xudata}.

\section*{Acknowledgments}
The authors thank the anonymous referee for suggestions.

\end{document}